\documentclass[11pt]{article}

\usepackage[margin=1in]{geometry}
\usepackage{amsmath,amssymb,amsfonts,bm}
\usepackage{graphicx}
\usepackage{booktabs}
\usepackage{xcolor}
\usepackage[colorlinks=true,linkcolor=blue,citecolor=blue,urlcolor=blue]{hyperref}

\title{Wave--Tide Locking in Thin Stellar Streams: A Phenomenological Mass Spectrometer for an Intermediate Ultralight Axion}

\author{Peter H. Tsang\thanks{Email: \texttt{phtsang@alum.mit.edu}}}

\date{\today}

\begin{document}
\maketitle

\begin{abstract}
We propose a phenomenological mass estimator for an intermediate ultralight axion dark-matter component using thin stellar streams. The central observation is an analytic relation linking three length scales in a fuzzy-dark-matter stream progenitor: the tidal radius of a bound dark core, its gravitational Bohr radius, and the axion de Broglie wavelength evaluated at the stripped-star velocity scale. If the stream width is $w=C_w r_{\rm t}$, with $C_w=O(1)$, then
\begin{equation}
    \frac{r_{\rm t}}{r_{\rm B}}
    =
    \frac{4\pi^2}{C_w^2}
    \left(\frac{w}{\lambda_{\rm dB}}\right)^2 .
\end{equation}
Thus the condition $\lambda_{\rm dB}\sim w$ automatically implies $r_{\rm t}/r_{\rm B}\sim25$--$40$: the dark wave core survives well inside the tidal boundary while the extended stellar envelope is stripped into a thin stream. This wave--tide locking gives a no-simulation axion-mass estimator,
\begin{equation}
    m_a \simeq
    2.69\times10^{-19}\,\mathrm{eV}
    \left(\frac{\sqrt{2}}{q_\kappa}\right)
    \left(\frac{R}{10\,\mathrm{kpc}}\right)
    \left(\frac{38\,\mathrm{pc}}{w}\right)^2
    \left(\frac{w}{\ell_{\rm ripple}}\right)
    \left(\frac{220\,\mathrm{km\,s^{-1}}}{v_c}\right),
\end{equation}
where $R$ is Galactocentric radius, $q_\kappa=\kappa/\Omega$, and $\ell_{\rm ripple}$ is an observed small-scale ripple or coherence length identified with $\lambda_{\rm dB}$. In the simplest locked case $\ell_{\rm ripple}\sim w$, a homogeneous first-pass set of narrow stream widths points to $m_a$ of a few $10^{-19}\,\mathrm{eV}$ and hidden-core masses of a few $10^3$--$10^4M_\odot$. This is not a detection claim; it is a falsifiable phenomenological test for an ultralight axion component that can be sharpened by homogeneous stream catalogs and Schr\"odinger--Poisson simulations.
\end{abstract}

\section{Introduction}

Cold dark matter successfully explains the large-scale structure of the Universe, but its small-scale phenomenology remains an active testing ground. Ultralight or fuzzy dark matter replaces classical particle behavior by coherent wave dynamics, producing de Broglie-scale pressure support, solitonic cores, and interference-like granularity in virialized halos \cite{Hu2000,Hui2017,Schive2014,Niemeyer2020}. Stellar streams are among the most sensitive probes of such subgalactic structure, because their small internal velocity dispersions preserve gravitational perturbations over long times \cite{Bonaca2025}.

From a particle-physics phenomenology viewpoint, the useful question is not only whether an ultralight axion can form structure, but whether astrophysical observables can be inverted into an axion mass estimator. The goal here is therefore to formulate a falsifiable low-energy dark-matter phenomenology observable, not to claim an astrophysical detection.

The purpose of this paper is analytic. We ask whether thin stellar streams can act as mass spectrometers for an intermediate ultralight axion species, without first requiring a full numerical simulation. The main result is a wave--tide locking theorem: if the stream width is comparable to the axion de Broglie wavelength, then the dark core is automatically well inside its tidal radius.

\section{Main results}

The paper has four main results. First, the de Broglie scale is the geometric bridge between the tidal radius and the compact core radius:
\begin{equation}
    \bar\lambda_{\rm dB}^2=r_{\rm t}r_{\rm B},
    \qquad
    \lambda_{\rm dB}^2=4\pi^2 r_{\rm t}r_{\rm B}.
\end{equation}
Second, if the physical stream width is $w=C_w r_{\rm t}$, then
\begin{equation}
    \frac{r_{\rm t}}{r_{\rm B}}
    =
    \frac{4\pi^2}{C_w^2}
    \left(\frac{w}{\lambda_{\rm dB}}\right)^2 .
\end{equation}
Thus $\lambda_{\rm dB}\sim w$ implies $r_{\rm t}/r_{\rm B}\sim25$--$40$ for Milky-Way-like streams.

Third, the locked relation gives an observable axion-mass estimator. In the simple flat-rotation case and for a ripple scale comparable to the width,
\begin{equation}
    m_a^{\rm lock}
    \simeq
    2.69\times10^{-19}\,\mathrm{eV}
    \left(\frac{R}{10\,\mathrm{kpc}}\right)
    \left(\frac{38\,\mathrm{pc}}{w}\right)^2.
\end{equation}
Clean streams should therefore approximately satisfy $w\propto R^{1/2}$ if they share a common locked axion mass.

Fourth, applying the estimator to a homogeneous first-pass set of narrow stream widths gives an intermediate ultralight scale of a few $10^{-19}\,\mathrm{eV}$ and hidden masses of a few $10^3$--$10^4\,M_\odot$.

\section{Assumptions and regime of validity}

The analytic model is intentionally minimal. We assume a localized fuzzy-dark-matter core of mass $M$ and axion mass $m_a$ moving in a smooth Galactic tidal field. The core is treated in the Newtonian Schr\"odinger--Poisson regime, so its characteristic radius is the gravitational Bohr radius. The surrounding stars are treated as collisionless tracers rather than as a self-gravitating stellar system. This is appropriate when the stellar mass in the visible stream progenitor is small compared with the hidden core mass.

We also assume that the stream segment being analyzed is a cold, narrow component whose width is set primarily by tidal stripping from the progenitor. Broad cocoons, fanning structures, dwarf-galaxy debris, and strongly perturbed stream segments are not clean applications of the estimator.

The host potential is approximated locally by an angular frequency $\Omega$ and epicyclic frequency $\kappa=q_\kappa\Omega$. The theorem does not require an exactly flat rotation curve; departures from flatness enter through the order-one factor $q_\kappa$. The uncertain convention relating stream width to tidal radius is collected into another order-one factor $C_w$.

Finally, identifying a ripple length with the axion de Broglie wavelength should be understood as a testable hypothesis, not as an assumption that every stream density variation is a wave-interference feature.

\section{Self-gravitating axion core and tidal radius}

Consider a localized fuzzy-dark-matter core of mass $M$ and axion mass $m_a$. In the Newtonian Schr\"odinger--Poisson regime its characteristic gravitational Bohr radius is
\begin{equation}
    r_{\rm B}=\frac{\hbar^2}{G M m_a^2}.
    \label{eq:bohr}
\end{equation}
For the intermediate mass $m_\mu=2.89\times10^{-19}\,\mathrm{eV}$ this becomes
\begin{equation}
    r_{\rm B}
    \simeq
    1.023\,\mathrm{pc}
    \left(\frac{10^4M_\odot}{M}\right)
    \left(\frac{2.89\times10^{-19}\,\mathrm{eV}}{m_a}\right)^2 .
    \label{eq:bohr_numeric}
\end{equation}

For an object orbiting at Galactocentric radius $R$ in a host with circular speed $v_c$ and angular frequency $\Omega=v_c/R$, write the tidal radius as
\begin{equation}
    r_{\rm t}
    =
    \left(\frac{G M}{\alpha_t \Omega^2}\right)^{1/3}
    =
    \left(\frac{G M R^2}{\alpha_t v_c^2}\right)^{1/3},
    \label{eq:rt_general}
\end{equation}
where $\alpha_t$ is an order-one tidal coefficient.

Taking $\alpha_t=3$ and $v_c=220\,\mathrm{km\,s^{-1}}$ gives
\begin{equation}
    r_{\rm t}
    \simeq
    30.9\,\mathrm{pc}
    \left(\frac{M}{10^4M_\odot}\right)^{1/3}
    \left(\frac{R}{10\,\mathrm{kpc}}\right)^{2/3}
    \left(\frac{220\,\mathrm{km\,s^{-1}}}{v_c}\right)^{2/3}.
    \label{eq:rt_numeric}
\end{equation}

\section{Stream width and hidden-core mass estimator}

Let $\sigma_{\rm strip}$ be the characteristic velocity scale of stars stripped near $r_{\rm t}$:
\begin{equation}
    \sigma_{\rm strip}^2\simeq \frac{G M}{r_{\rm t}} .
\end{equation}
The stream width is estimated by epicyclic spreading,
\begin{equation}
    w\sim \frac{\sigma_{\rm strip}}{\kappa},
\end{equation}
where $\kappa=q_\kappa\Omega$ is the epicyclic frequency.

Using $G M=\alpha_t\Omega^2 r_{\rm t}^3$, one obtains
\begin{equation}
    w=C_w r_{\rm t},
    \qquad
    C_w\equiv \frac{\sqrt{\alpha_t}}{q_\kappa}.
    \label{eq:cw}
\end{equation}
For $\alpha_t=2$--$3$ and a flat Milky-Way-like rotation curve, $C_w\simeq1.0$--$1.2$.

Using the fiducial convention $\alpha_t=3$ and $q_\kappa=\sqrt2$, one has $C_w=\sqrt{3/2}$ and
\begin{equation}
    w
    \simeq
    37.8\,\mathrm{pc}
    \left(\frac{M}{10^4M_\odot}\right)^{1/3}
    \left(\frac{R}{10\,\mathrm{kpc}}\right)^{2/3}
    \left(\frac{220\,\mathrm{km\,s^{-1}}}{v_c}\right)^{2/3}.
    \label{eq:w_numeric}
\end{equation}
Inverting,
\begin{equation}
    M_w
    \simeq
    10^4M_\odot
    \left(\frac{w}{38\,\mathrm{pc}}\right)^3
    \left(\frac{10\,\mathrm{kpc}}{R}\right)^2
    \left(\frac{v_c}{220\,\mathrm{km\,s^{-1}}}\right)^2.
    \label{eq:mass_estimator}
\end{equation}

\section{Wave--tide locking theorem}

The reduced de Broglie length and full de Broglie wavelength are
\begin{equation}
    \bar\lambda_{\rm dB}
    =
    \frac{\hbar}{m_a\sigma_{\rm strip}},
    \qquad
    \lambda_{\rm dB}
    =
    \frac{h}{m_a\sigma_{\rm strip}}
    =
    2\pi\bar\lambda_{\rm dB}.
\end{equation}
Using $\sigma_{\rm strip}^2=GM/r_{\rm t}$ and Eq.~\eqref{eq:bohr},
\begin{equation}
    \bar\lambda_{\rm dB}^2
    =
    \frac{\hbar^2}{m_a^2\sigma_{\rm strip}^2}
    =
    \frac{\hbar^2}{m_a^2}
    \frac{r_{\rm t}}{GM}
    =
    r_{\rm t}r_{\rm B}.
    \label{eq:reduced_geom_mean}
\end{equation}
Equivalently,
\begin{equation}
    \lambda_{\rm dB}^2
    =
    4\pi^2 r_{\rm t}r_{\rm B}.
    \label{eq:full_geom_mean}
\end{equation}

Combining Eq.~\eqref{eq:full_geom_mean} with $w=C_w r_{\rm t}$ gives
\begin{equation}
    \boxed{
    \frac{r_{\rm t}}{r_{\rm B}}
    =
    \frac{4\pi^2}{C_w^2}
    \left(\frac{w}{\lambda_{\rm dB}}\right)^2 .}
    \label{eq:locking_general}
\end{equation}
For Milky-Way-like streams, $C_w\simeq1.0$--$1.2$, so
\begin{equation}
    \frac{r_{\rm t}}{r_{\rm B}}
    \simeq
    (25\text{--}40)
    \left(\frac{w}{\lambda_{\rm dB}}\right)^2 .
    \label{eq:locking_numeric}
\end{equation}
Therefore, if $\lambda_{\rm dB}\sim w$, then $r_{\rm t}/r_{\rm B}\sim25$--$40$.

\section{Observable axion-mass estimator}

Because $\sigma_{\rm strip}=\kappa w=q_\kappa v_c w/R$, the axion mass can be inferred from a measured ripple scale $\ell_{\rm ripple}$ identified with $\lambda_{\rm dB}$:
\begin{equation}
    m_a
    =
    \frac{h}{\ell_{\rm ripple}\sigma_{\rm strip}}
    =
    \frac{hR}{q_\kappa v_c w\ell_{\rm ripple}}.
\end{equation}
Writing $\ell_{\rm ripple}=\Xi w$ gives
\begin{equation}
    \boxed{
    m_a
    \simeq
    2.69\times10^{-19}\,\mathrm{eV}
    \left(\frac{\sqrt2}{q_\kappa}\right)
    \left(\frac{R}{10\,\mathrm{kpc}}\right)
    \left(\frac{38\,\mathrm{pc}}{w}\right)^2
    \left(\frac{1}{\Xi}\right)
    \left(\frac{220\,\mathrm{km\,s^{-1}}}{v_c}\right).}
    \label{eq:ma_estimator}
\end{equation}
Equivalently,
\begin{equation}
    m_a
    \simeq
    2.69\times10^{-19}\,\mathrm{eV}
    \left(\frac{\sqrt2}{q_\kappa}\right)
    \left(\frac{R}{10\,\mathrm{kpc}}\right)
    \left(\frac{38\,\mathrm{pc}}{w}\right)^2
    \left(\frac{w}{\ell_{\rm ripple}}\right)
    \left(\frac{220\,\mathrm{km\,s^{-1}}}{v_c}\right).
\end{equation}

\section{Representative thin-stream sample}

Table~\ref{tab:candidates} applies the analytic estimators to a representative set of narrow stream components. The values are intended as a first regime map, not a final homogeneous catalog. The first-pass physical widths are taken from the homogeneous modeling of Patrick et al.~\cite{Patrick2022}. A final catalog analysis should replace the effective radius used here by an orbit-averaged Galactocentric radius.

\begin{table}[ht]
\centering
\caption{Homogeneous-width first-pass stream table. The Pal 5 leading and trailing tails are kept separate because Patrick et al.~\cite{Patrick2022} model them separately.}
\label{tab:candidates}
\begin{tabular}{lcccccc}
\toprule
Stream & $R_{\rm eff}$ [kpc] & $w$ [pc] & $M_w$ [$M_\odot$] & $r_B$ [pc] & $r_t/r_B$ & $m_a^{\rm lock}$ [eV] \\
\midrule
Phoenix & 18.1 & 41 & $3.8\times10^3$ & 2.67 & 12.5 & $4.2\times10^{-19}$ \\
ATLAS & 21.1 & 59 & $8.4\times10^3$ & 1.22 & 39.6 & $2.4\times10^{-19}$ \\
Tucana III & 23.6 & 70 & $1.1\times10^4$ & 0.91 & 62.7 & $1.9\times10^{-19}$ \\
GD-1 & 8.6 & 30 & $6.7\times10^3$ & 1.54 & 15.9 & $3.7\times10^{-19}$ \\
Pal 5 leading & 19.0 & 38 & $2.8\times10^3$ & 3.69 & 8.4 & $5.1\times10^{-19}$ \\
Pal 5 trailing & 19.0 & 45 & $4.6\times10^3$ & 2.22 & 16.5 & $3.6\times10^{-19}$ \\
\bottomrule
\end{tabular}
\end{table}

Using this homogeneous-width first pass, the listed components give a log-mean locked axion mass of approximately $3.3\times10^{-19}\,\mathrm{eV}$, with scatter about $0.15$ dex. They also give a log-mean width-inferred hidden mass of approximately $5.6\times10^3M_\odot$, with scatter about $0.21$ dex.

A key remaining systematic is the replacement of the true orbit-averaged Galactocentric radius by the effective radius used in Table~\ref{tab:candidates}. The scalings are
\begin{equation}
    M_w\propto R^{-2},
    \qquad
    m_a^{\rm lock}\propto R.
\end{equation}
Thus a fractional error $R\rightarrow f_RR$ sends $M_w\rightarrow f_R^{-2}M_w$ and $m_a^{\rm lock}\rightarrow f_Rm_a^{\rm lock}$.

\begin{figure}[ht]
\centering
\includegraphics[width=0.82\textwidth]{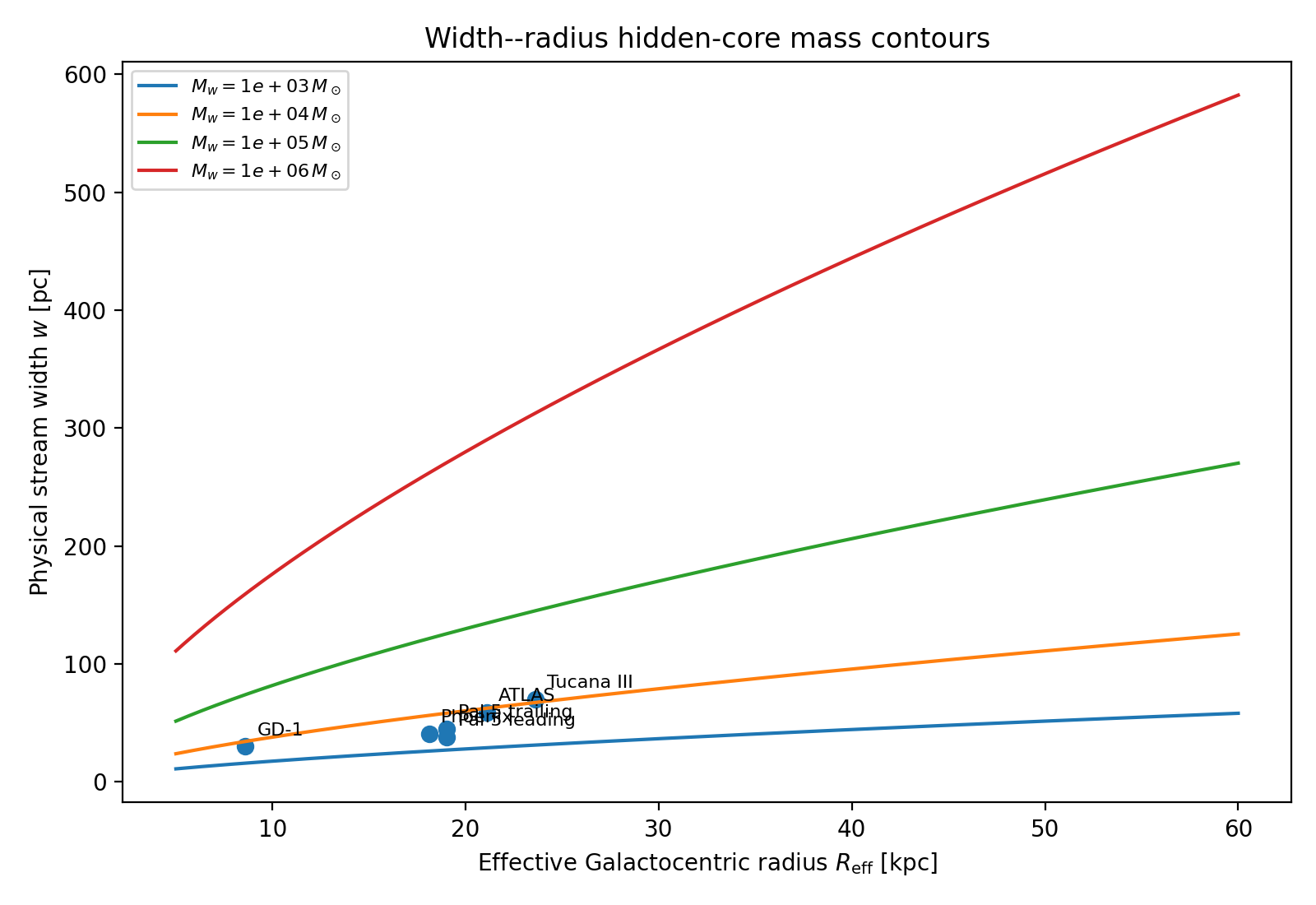}
\caption{Width--radius hidden-core mass contours. The curves show the analytic estimator for hidden-core mass as a function of stream width and Galactocentric radius. Points show the homogeneous-width first-pass sample.}
\label{fig:width_radius}
\end{figure}

\begin{figure}[ht]
\centering
\includegraphics[width=0.82\textwidth]{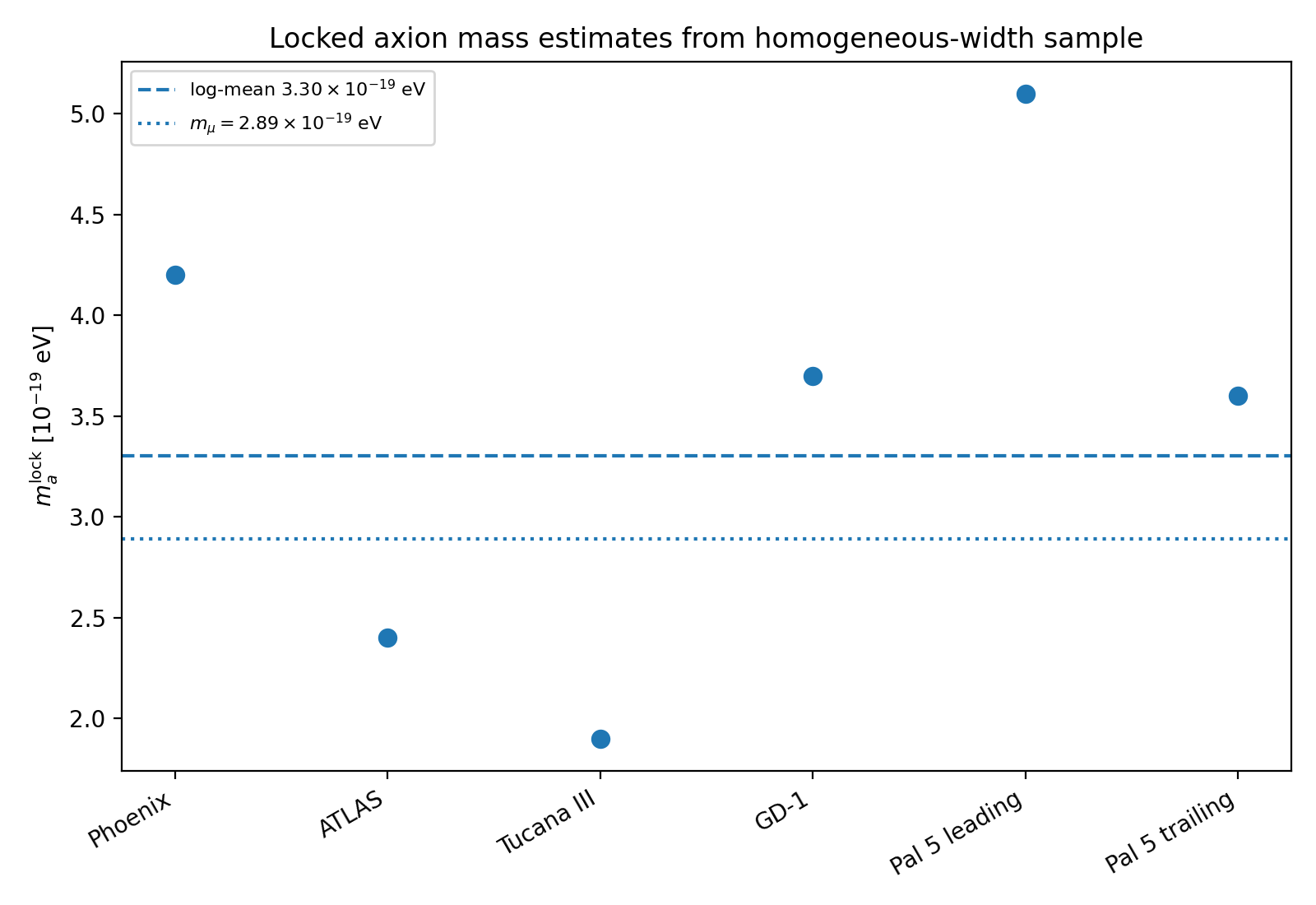}
\caption{Locked axion-mass estimates for the homogeneous-width first-pass sample. The dashed line shows the log-mean of the sample and the dotted line marks the proposed intermediate species.}
\label{fig:locked_mass}
\end{figure}

\begin{figure}[ht]
\centering
\includegraphics[width=0.82\textwidth]{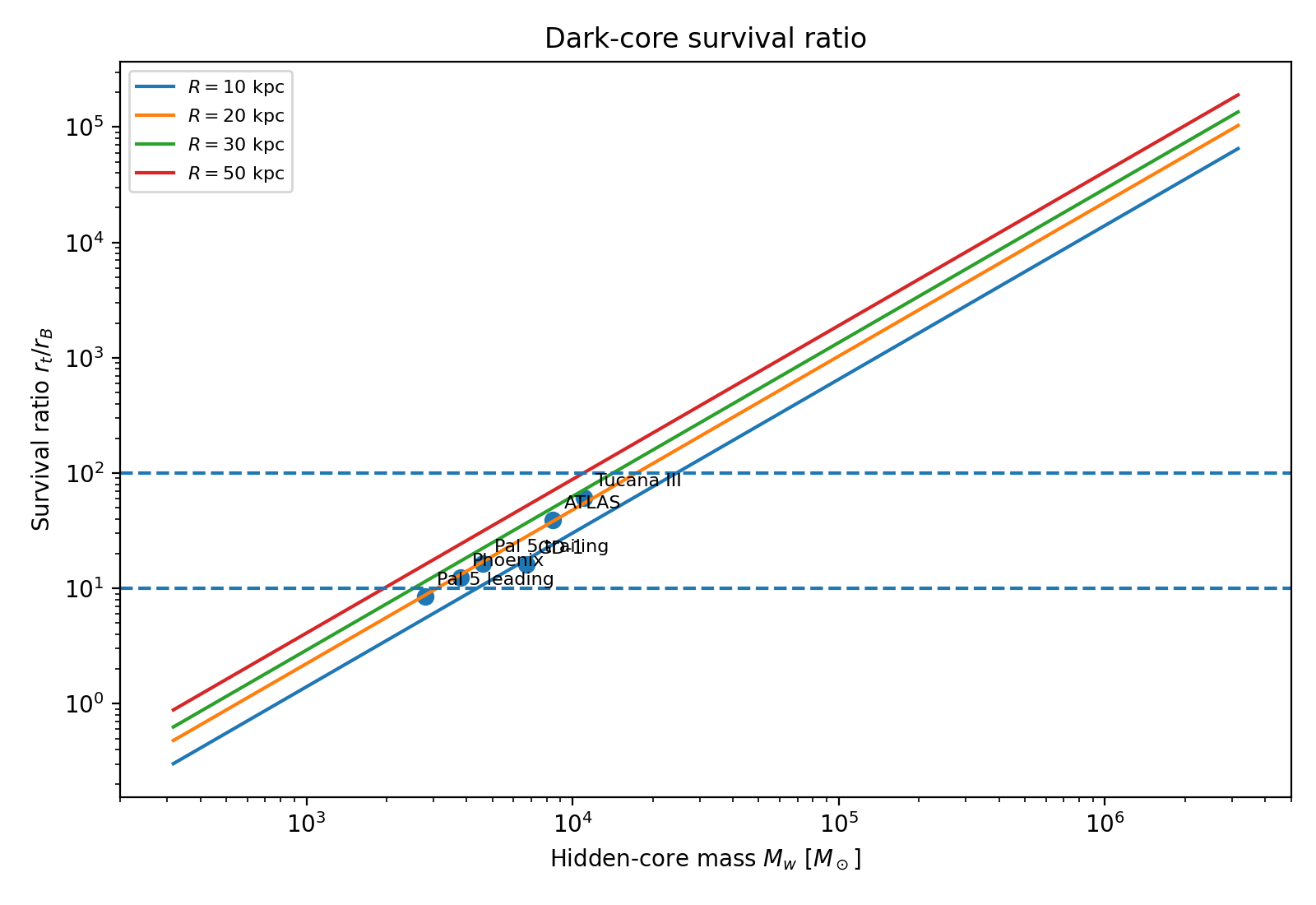}
\caption{Dark-core survival ratio as a function of hidden-core mass. The horizontal dashed lines mark the approximate clean survival window.}
\label{fig:survival_ratio}
\end{figure}

\section{Comparison with ordinary stream interpretations}

Ordinary tidal-stream physics is an important null hypothesis. A globular cluster or dwarf galaxy on an eccentric Galactic orbit can produce a narrow stream with width set by the progenitor tidal radius. Classical dark-matter subhalos can also create gaps, kinks, and density variations. Therefore, stream width alone cannot prove an axion interpretation.

The distinctive feature of the present model is not merely that a hidden mass can set a tidal width. It is the linked scale system
\begin{equation}
    M\sim10^4M_\odot,\qquad
    r_B\sim1\,\mathrm{pc},\qquad
    r_t\sim w\sim\lambda_{\rm dB}\sim\mathrm{tens\ of\ pc}.
\end{equation}
A classical globular-cluster stream may reproduce the tidal-width part of this system, but it does not predict a de Broglie scale tied to the stream width.

\section{Relation to a three-species axion hierarchy}

The three-species interpretation used here is motivated by the multicomponent axionic dark-sector scenario of Meinert and Hofmann \cite{MeinertHofmann2021}, in which fuzzy dark matter is treated as a three-component axionic fluid and the lightest axion mass is extracted from low-surface-brightness galaxy rotation-curve phenomenology. Related SU(2) Yang--Mills thermodynamic arguments for axion-mass generation have been developed by Hofmann, Meinert, and Antonov \cite{HofmannMeinertAntonov2024}. The present paper does not assume the full cosmological model of these works. Instead, it asks whether the intermediate mass scale can be tested independently through a local stream observable: wave--tide locking in cold, narrow stellar streams.

The intermediate species is selected by the stream problem. The lightest species,
\begin{equation}
    m_e\sim6.75\times10^{-24}\,\mathrm{eV},
\end{equation}
would give $10^4M_\odot$ objects with radii far too diffuse to act as compact stream cores. It is instead naturally associated with kpc-scale galactic backgrounds. The heaviest species,
\begin{equation}
    m_\tau\sim8.17\times10^{-17}\,\mathrm{eV},
\end{equation}
would be much too compact for the clean thin-stream mechanism and instead approaches a relativistic compact-object boundary.

Comparing the Bohr radius to the Schwarzschild radius gives
\begin{equation}
    M_{\rm rel}(m_a)=\frac{\hbar c}{\sqrt2 Gm_a}.
\end{equation}
For $m_\tau$ this yields
\begin{equation}
    M_{\rm rel,\tau}\simeq1.16\times10^6M_\odot.
\end{equation}
This gives a clean division of roles:
\begin{equation}
    e:\ \mathrm{kpc\ galactic\ background},\qquad
    \mu:\ 10^4M_\odot\ \mathrm{thin\ stream\ cores},\qquad
    \tau:\ 10^6M_\odot\ \mathrm{compact\ boundary}.
\end{equation}

\section{Simulation target}

Although the central theorem is analytic, simulations are the next step. For the fiducial intermediate species and stream-core mass,
\begin{equation}
    M_\mu=10^4M_\odot,\qquad
    r_B\simeq1.02\,\mathrm{pc},\qquad
    v_B\simeq6.5\,\mathrm{km\,s^{-1}},\qquad
    t_B\simeq0.154\,\mathrm{Myr}.
\end{equation}
At $R\sim10$--$25\,\mathrm{kpc}$, one expects
\begin{equation}
    r_t\sim30\text{--}60\,\mathrm{pc},
    \qquad
    w\sim40\text{--}80\,\mathrm{pc}.
\end{equation}
A minimal reduced simulation consists of a resolved Schr\"odinger--Poisson $\mu$-core, a smooth harmonic background potential from the lightest species, an external Milky-Way tidal field, and cold stellar tracers with scale radius $a_\star\sim20$--$50\,\mathrm{pc}$.

\section{Discussion}

The main result of this paper is not that axions have been detected. Rather, it is that clean thin streams offer an analytic overconstraint. Width estimates a hidden core mass, small-scale ripples can estimate a de Broglie wavelength, and the wave--tide locking theorem connects both to dark-core survival.

The most important limitations are:
\begin{enumerate}
    \item The stream table uses representative first-pass values. A final analysis must homogenize widths, radii, distances, and component definitions.
    \item The ripple scale $\ell_{\rm ripple}$ should not be assumed to be de Broglie interference without phase-space evidence.
    \item Ordinary globular-cluster streams can reproduce tidal widths. The distinct fuzzy signature is the linked scale set.
    \item The $\tau$ compact-object sector requires relativistic treatment and should not be confused with the Newtonian $\mu$ stream-core sector.
\end{enumerate}

\section{Conclusions}

We have derived a wave--tide locking relation for fuzzy-dark-matter stream progenitors:
\begin{equation}
    \frac{r_t}{r_B}
    =
    \frac{4\pi^2}{C_w^2}
    \left(\frac{w}{\lambda_{\rm dB}}\right)^2 .
\end{equation}
For Milky-Way-like streams, $C_w\sim1$--$1.2$, so $\lambda_{\rm dB}\sim w$ implies $r_t/r_B\sim25$--$40$. This is precisely the regime where a compact dark axion core survives while an extended stellar envelope is stripped into a thin stream.

The resulting observable estimator maps clean thin streams to an axion mass near a few $10^{-19}\,\mathrm{eV}$ and to hidden-core masses near $10^4M_\odot$. Thin stellar streams may therefore function as analytic mass spectrometers for an intermediate ultralight axion component.

\appendix

\section{Conventions and numerical normalizations}

The reduced de Broglie length uses $\hbar$, while the full de Broglie wavelength uses $h$. The Bohr radius uses $\hbar^2$. This gives
\begin{equation}
    \bar\lambda_{\rm dB}^2=r_t r_B,
    \qquad
    \lambda_{\rm dB}^2=4\pi^2 r_t r_B.
\end{equation}
Writing $w=C_w r_t$ gives
\begin{equation}
    \frac{r_t}{r_B}
    =
    \frac{4\pi^2}{C_w^2}
    \left(\frac{w}{\lambda_{\rm dB}}\right)^2.
\end{equation}

\section{Claims and limitations}

The result in this paper is an analytic consistency test, not a detection claim. The present draft does not claim that all stellar streams arise from axion cores, that all dark matter is in compact $10^4M_\odot$ objects, or that the representative stream table is a final catalog measurement.

\section{Data and code availability}

No new observational data are introduced in this analytic draft. The first-pass stream widths are taken from the homogeneous modeling of Patrick et al.~\cite{Patrick2022}. Future versions should use public stream-track resources such as \texttt{galstreams} \cite{Mateu2023} to compute orbit-averaged Galactocentric radii for the same stream segments used in the width measurements.

\section*{Acknowledgments}

The author thanks Ralf Hofmann and Janning Meinert for helpful discussions, encouragement, and support.

\section*{Dedication}

This paper is dedicated to the memory of Herbert Martin Fried and Chung-I Tan, whose guidance, insight, encouragement, and example shaped the author's path in theoretical physics. Their ideas and generosity continue to inspire this work.

\bibliographystyle{unsrt}
\bibliography{wave_tide_locking_refs}

\end{document}